\documentclass[11pt,aps,prd,nofootinbib,floats,floatfix,preprintnumbers,groupedaddress]{revtex4}
\usepackage{comment}
\usepackage{graphicx}
\usepackage{amsmath, mathrsfs}
\usepackage{float}
\usepackage{subfigure}
\usepackage[usenames]{color}
\usepackage{amssymb}
\usepackage{bbm}
\usepackage{csquotes}
\usepackage{color}
\usepackage{psfrag}
\usepackage{graphicx}
\usepackage{hyperref}
\begin{document}

\title{Criticality Quenching and Microstructure of Quintessence-AdS Black Holes}

\author{Apurba Tiwari$^1$, Randeep Kaur$^2$, Javed Khan Bhutto$^3$, Thafasalijyas Vayalpurayil$^4$ and Mohammed Sayeeduddin Habeeb$^5$}
\email{apurba.e17565@cumail.in} \email{randeep.e13927@cumail.in} \email{jbhutto@kku.edu.sa} 
\email{salman@kku.edu.sa} 
\email{mshabeeb@kku.edu.sa}

\affiliation{$^{1,2}$Department of Physics, University Institute of Sciences, Chandigarh University, Mohali, Punjab 140413, India\\
$^3$ Department of Electrical Engineering, College of Engineering, King Khalid University, Abha, Saudi Arabia\\
$^{4,5}$Electrical Engineering Department, Computer Engineering Section, College of Engineering, King Khalid University, Abha, KSA} 

	\vspace{0.5cm}
	
\begin{abstract}
In this work, we investigate the thermodynamic geometry of Reissner-Nordström Anti-de Sitter (RN-AdS) black holes with quintessence in the grand canonical ensemble. The analysis employs the Ruppeiner curvature scalar to elucidate the microscopic interactions and critical phenomena in the extended phase space. Divergence of the scalar curvature signal phase transitions, while its sign characterizes the nature of the underlying interactions; negative for attractive and positive for repulsive type interactions. The analysis reveals that attractive interactions dominate at low electric potentials, whereas repulsive interactions prevail at higher potentials unlike the usual charged black holes. Finally, the interaction strength is fairly constant during the phase transition, providing a greater understanding of the quintessence influenced microscopic attributes of RN-AdS black holes.
\end{abstract}

\maketitle
\section{Introduction}

It is now a well-established fact that the present universe is undergoing accelerated expansion, a phenomenon strongly supported by various independent astronomical observations, including Type Ia supernovae luminosity distances, cosmic microwave background (CMB) anisotropies, baryon acoustic oscillations (BAO), and large-scale structure surveys~\cite{Filippenko:2003pq,SupernovaSearchTeam:1998fmf}. This late-time cosmic acceleration is attributed to an exotic component possessing large negative pressure, commonly referred to as dark energy, which contributes nearly $70\%$ of the total energy density of the universe~\cite{Branch:1992rv,Caldwell:1997ii}. Among several theoretical models proposed to describe dark energy, the concept of quintessence has emerged as one of the most viable candidates. Quintessence is characterized by a dynamical scalar field $\phi(t)$ slowly evolving under a self-interacting potential $V(\phi)$, producing an effective equation of state parameter satisfying $-1 < \omega_q < -1/3$~\cite{Kiselev}. Unlike the cosmological constant~\cite{Carroll:1991mt}, which corresponds to $\omega = -1$, quintessence exhibits a time-varying energy density that can influence both cosmological dynamics and local gravitational configurations. The spacetime structure surrounding compact objects, like black holes, can be drastically altered by quintessence. It is possible to model the quintessential field as a fluid that permeates spacetime~\cite{Azreg}, changing the asymptotic and near-horizon geometries. For a black hole enveloped by such a quintessence field, Kiselev~\cite{Kiselev} originally offered a spherically symmetric solution in which the stress-energy tensor fulfills a linear relation between its temporal and spatial components. Since then, a great deal of research has been done on the physical and thermodynamic consequences of these structures~\cite{Visser,quint1,Ghosh:2017cuq,Ghosh:2016ddh,Ghaderi:2016ttd,Ghaderi:2016dpi,Toledo,quint2}. In such circumstances, the black hole's thermodynamic properties, stability, and horizon structure are influenced by the intensity and form of the dark energy environment surrounding it, which is determined by the key parameter $\omega_q$ and normalizing factor $\alpha$.

In this paper, we explore the four-dimensional Reissner-Nordström Anti-de Sitter (RN-AdS) black hole with quintessence in the grand canonical ensemble with fixed electric potential $\Phi$. Compared to the canonical ensemble, the grand canonical description offers a more broader thermodynamic framework and is physically appealing since it naturally includes charge interaction among the black hole and the surrounding space.   Using the effective formalism of Ruppeiner geometry, the main goal of this work is to comprehend how the existence of quintessence alters the thermodynamic phase structure and microscopic interaction patterns of RN-AdS black holes. The extended phase space approach, which treats the cosmological constant $\Lambda$ as a thermodynamic pressure via the relation~\cite{Kastor,Sahni:1999gb},
\begin{equation}\label{cosmological constant}
P = -\frac{\Lambda}{8\pi}
\end{equation}
has been used to reexamine the thermodynamics of AdS black holes all throughout the past decade. In accordance to the generalized Smarr relation~\cite{Cvetic:2010jb} and the extended first law, the black hole mass $M$ is reinterpreted as the system's enthalpy in this formalism. This reinterpretation has led to an interesting comparison between the Van der Waals liquid-gas system and charged AdS black holes. RN-AdS black holes show first-order phase transitions similar to those in standard thermodynamic fluids, according to early research by Chamblin \textit{et al.}~\cite{Chamblin:1999tk,Chamblin:1999hg}. Later research by Kubizňák and Mann~\cite{Kubiznak:2012wp} formalized this analogy by identifying the critical points in the $P$-$V$ diagram. The aforementioned developments made it possible to understand black hole phase transitions in terms of intermolecular interactions and microscopic degrees of freedom. Due to its relation to the event horizon's area, black hole entropy suggests a strong connection between the underlying microscopic structure of spacetime and the macroscopic thermodynamic quantities~\cite{Modak:2025gvp}. Although the exact nature of these microstructures is still unknown, a number of theories attempt to explain exactly how they are evolved. The fuzzball paradigm~\cite{Bianchi:2020bxa} in string theory states that each black hole microstate is linked to a smooth, horizonless geometry that mimics the black hole's thermodynamic properties. Alternatively, the Cardy formula~\cite{DavoodSadatian:2017vsy} can be used to derive the entropy–area connection from conformal field theory considerations, supporting the holographic interpretation of black hole thermodynamics. The surface gravity $\kappa$ and the Hawking temperature $T_h$ can be determined as,
 \begin{equation}
 T_h = \frac{\kappa}{2\pi}
 \end{equation}
 demonstrates that the black hole horizon has microscopic degrees of freedom that collectively encode thermal behavior, strengthening the thermodynamic comparison~\cite{Hawking:1982dh,Bekenstein:1973ur,Hawking:1976de}. The equipartition theorem~\cite{Padmanabhan} suggests that the black hole microstates can be thought of as effective “molecules” similar to those in a regular thermodynamic fluid as each degree of freedom contributes equally to the total energy.

Extensive studies have shown that Ruppeiner geometry provides valuable insights into the microscopic nature of black holes and their phase transitions~\cite{Sahay,Banerjee:2010da,Niu,Dehyadegari,Moumni,Kumara:2019xgt,Kumara:2020,Kumara:2020ucr,Wei:2019uqg,Wei2019b,AR,Xu,Ghosh,Wei,BTZ}. To investigate the statistical interaction among the black holes microstructures, the Ruppeiner thermodynamic geometry~\cite{Ruppeiner,Ruppeiner2,Ruppeiner3,RuppeinerRMP,Weinhold,Singh:2020tkf,Singh:2023hit,NooriGashti:2025sau,Anand:2025vvg,Anand:2026abc,Singh:2023ufh,Singh:2024msw,Singh:2025ueu,Anand:2025iib,Anand:2025mlc,Anand:2025vfj,Janyszek1990,Janyszek:1989zz,interactions,Bardeen:1973gs,Tiwari:2025ddq} offers an elegant geometric approach. In this formalism, the Ruppeiner metric is defined as the negative Hessian of the entropy with respect to the extensive thermodynamic variables, providing a Riemannian structure to the equilibrium thermodynamic state space. The associated scalar curvature $R$ encodes information about the underlying interactions: $R < 0$ indicates that attractive interactions dominate, $R > 0$ corresponds to repulsive interactions, and $R = 0$ signifies non-interacting microstructures~\cite{Janyszek:1989zz,Janyszek1990,interactions,NooriGashti:2025sau,Anand:2025vvg,Anand:2026abc}. The magnitude of $R$ is also related to the correlation length near criticality, diverging at phase transition points where macroscopic fluctuations become long-ranged. In this work, we construct the Ruppeiner geometry for RN-AdS black holes surrounded by quintessence in the grand canonical ensemble and examine the behavior of the normalized scalar curvature across varying electric potentials. In accordance with the results we obtained, quintessence significantly affects the curvature structure, resulting in different interaction regimes. In particular, for lower electric potential values, the Ruppeiner curvature is shown to be negative, suggesting that attractive interactions predominate among the microstructures. On the other hand, the curvature turns positive at greater electric potentials, indicating the appearance of repulsive interactions. Near the phase transition point, the curvature stays about constant, indicating that the nature of microscopic interactions does not change suddenly throughout the transition. The results obtained strengthen our understanding of the statistical mechanical interpretation of RN-AdS black holes within the context of thermodynamic geometry and demonstrate the complex function of quintessence in influencing their microscopic behavior and phase structure.
  
  \noindent
  \textbf{Motivation:} 
  The motivation behind this study is twofold. Firstly, black holes surrounded by quintessence provide an excellent theoretical laboratory to investigate how dark energy influences both the macroscopic and microscopic properties of gravitational systems. Despite extensive literature on black hole thermodynamics, the microscopic interpretation of thermodynamic quantities; especially in the presence of quintessence, remains largely unexplored. Understanding the influence of quintessence on black hole microstructures is essential for establishing a possible connection between cosmological dark energy and horizon-scale physics. Secondly, the adoption of the grand canonical ensemble framework, where the electric potential $\Phi$ is fixed and the charge $Q$ is allowed to fluctuate, offers a more physically realistic scenario for AdS black holes interacting with an external reservoir. This ensemble not only generalizes the canonical description but also enables one to capture the effects of charge exchange and potential fluctuations, which are particularly relevant in the presence of a surrounding quintessential field. 
  
 The Ruppeiner curvature behavior and the thermodynamic phase structure are drastically changed by changes in the electric potential in the grand canonical ensemble. The grand canonical ensemble has a more dynamic and rich structure compared to the canonical ensemble, which has a constant charge. The sign of the Ruppeiner scalar from negative to positive changes as the electric potential increases, denotes a transition from attractive to repulsive interactions dominance among the microstructures of the black holes. Thus the surrounding quintessence field not only affects the thermodynamic stability but also reshapes the microscopic interaction within the black hole system. This implies that a deeper and more complete understanding of the connection between dark energy effects, phase transitions, and microstructural interactions from a thermodynamic geometric perspective can be acquired by analyzing RN-AdS black holes surrounded by quintessence in the grand canonical ensemble.

The organization of this paper is as follows. In Sec.~\ref{Thermodynamics}, we revisit the thermodynamic properties of the Reissner-Nordström AdS (RN-AdS) black hole with quintessence within the framework of the extended phase space. Subsection~\ref{PV} is devoted to examining the influence of the quintessence field on the $P$-$v$ critical behavior of the RN-AdS black holes and further in Sec.~\ref{Grand criticality}, we analyze the thermodynamics and criticality in the grand canonical ensemble which is not present in literature yet. In Sec.~\ref{Ruppeiner geometry}, we construct the Ruppeiner thermodynamic geometry and compute the corresponding normalized Ruppeiner curvature scalar to probe the microscopic interactions underlying the black hole system. Finally, Sec.~\ref{Remarks} presents the concluding remarks and discussions. Throughout this work, we employ natural units with $\hbar = c = k_{B} = G = 1$.

\section{Review of Thermodynamics of RN-AdS Black Holes with Quintessence}\label{Thermodynamics}
\begin{figure}[h!]
	\begin{center}
		\includegraphics[width=3.5in]{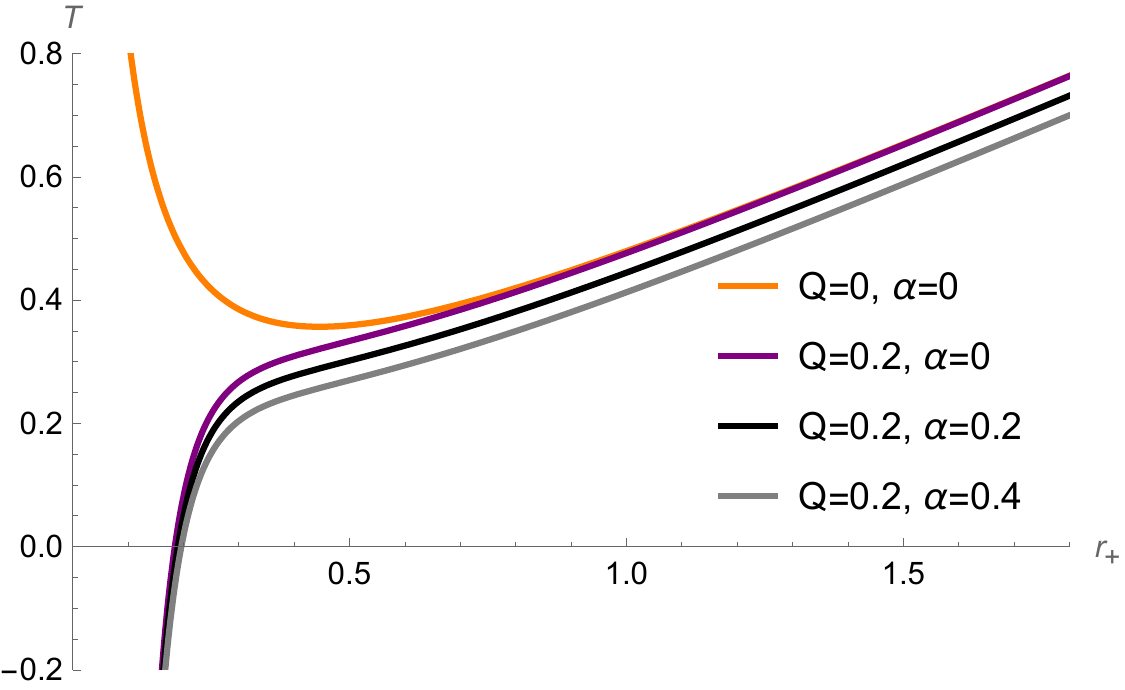} 
	\end{center}
	\caption{The behavior of black hole temperature $T$ with horizon radius $r_{+}$ for fixed values of $\omega_q=-2/3$ and pressure $P=0.2$.}
	\label{T_vs_r}
\end{figure}

In this section, we give a brief overview of the thermodynamics of RN-AdS black hole in four dimensions with quintessence. The corresponding metric is written as,
\begin{equation}
	dS^2 = f(r) dt^2 - f(r)^{-1} dr^2 - r^2 d\Omega_{2}^2,\label{1}
\end{equation}
where the lapse function takes the form,
\begin{equation}\label{lapse function}
	f(r) = 1 - \frac{2M}{r} + \frac{Q^2}{r^2} - \frac{\alpha}{r^{3\omega_q+1}} - \frac{\Lambda r^2}{3},
\end{equation}
and $\omega_q$ is the quintessential state parameter with the range $-1 < \omega_q < -1/3$. The normalization constant $\alpha$ characterizes the density of the quintessence field, which is given by
\begin{equation}
	\rho_q = -\frac{\alpha}{2}\frac{3\omega_q}{r^{3(\omega_q+1)}}.\label{3}
\end{equation}
Kiselev \cite{Kiselev} was the first to suggest the metric in Eq.~(\ref{lapse function}) as a solution to the Einstein field equations for a black hole surrounded by a quintessential matter field. In order to replicate the repulsive action that causes the cosmic acceleration seen in Type~Ia supernovae \cite{Filippenko:2003pq,SupernovaSearchTeam:1998fmf,SupernovaCosmologyProject:1997czu,SupernovaCosmologyProject:1998vns}, the crucial term alters the asymptotic structure of spacetime and adds a negative pressure component. The largest root of $f(r)|_{r=r_+} = 0$ yields the event horizon radius $r_+$. Utilzing Eq.~(\ref{cosmological constant}) and Eq.~(\ref{lapse function}), the corresponding mass (or equivalenty the enthalpy) of the black hole can be computed as,
\begin{equation}\label{Mass}
	M = \frac{r_+}{2}\left(1 + \frac{Q^2}{r_+^2} - \frac{\alpha}{r_+^{3\omega_q+1}} + \frac{8\pi P r_+^2}{3}\right).\
\end{equation}
The Hawking temperature, associated with the surface gravity at the horizon, is expressed as
\begin{equation}\label{Hawking temp}
	T = \frac{f'(r_+)}{4\pi} = \frac{1}{4\pi}\left(\frac{1}{r_+}-\frac{Q^2}{r_+^3}+\frac{3\alpha \omega_q}{ r_+^{2+3\omega_q}}+8\pi P r_+\right).
\end{equation}
Fig.~\ref{T_vs_r} illustrates how the presence of quintessence alters the black hole temperature and affects its stability by introducing an extra term proportional to $\alpha \omega_q$. This is demonstrated by equation~(\ref{Hawking temp}). When $\omega_q < 0$, this term lowers the temperature, suggesting that the surrounding dark energy field is cooling it down.
The Bekenstein-Hawking entropy  and the thermodynamic volume of the black hole can be evaluated using the standard thermodynamic definition given respectively as,
\begin{equation}\label{entropy}
	S = \int^{r_+}_0 \frac{1}{T}\left(\frac{\partial M}{\partial r_+}\right) dr_+ = \pi r_+^2, \quad V = \left(\frac{\partial M}{\partial P}\right)_{S,Q} = \frac{4\pi r_+^3}{3}.
\end{equation}
Consistent with other studies in extended thermodynamic frameworks, the entropy preserves its area law form, demonstrating that the existence of quintessence does not change the fundamental entropy-area relationship \cite{Kubiznak:2012wp}. The black hole's geometric volume and the equivalent thermodynamic volume are the same. This demonstrates that adding quintessence alters the temperature and mass relations but not the entropy or volume relations. According to the Smarr relation, the first law of black hole thermodynamics in the extended phase space \cite{Bardeen:1973gs,Kastor:2009wy} can also be expressed as follows: \begin{equation}
	dM = T dS + \Phi dQ + V dP + \mathcal{A} d\alpha,\label{first law}
	\end{equation}
where $\mathcal{A}$ is the quantity conjugate to the parameter $\alpha$. Utilizing, Eqs.~(\ref{Mass}) and (\ref{first law}), one can obtain the conjugate quantities,
\begin{eqnarray}
	\Phi = \left(\frac{\partial M}{\partial Q}\right)_{S,P,\alpha} = \frac{Q}{r_+}, \quad
	\mathcal{A} = \left(\frac{\partial M}{\partial \alpha}\right)_{S,Q,P} = -\frac{1}{2r_+^{3\omega_q}}.
\end{eqnarray}
Finally, the generalized Smarr relation reads,
\begin{equation}
	M = 2TS + \Phi Q - 2PV + (1 + 3\omega_q)\mathcal{A}\alpha.\label{smarr}
\end{equation}
Equation~(\ref{smarr}) shows that the critical parameter $\alpha$ introduces a term to the Smarr formula, which represents the black hole-dark energy field interaction. The black hole system's phase structure and microscopic interaction picture are affected by this extra degree of freedom, which alters the scaling behavior of thermodynamic quantities \cite{Azreg-Ainou:2014pra,Ayon-Beato:1998hmi}.

 % % % % % % % % % % % % % % % % %

 \subsection{Thermodynamic criticality of RN-AdS black holes with quintessence}\label{PV}
\begin{figure}[h!]
	\centering
	\includegraphics[width=3.5in]{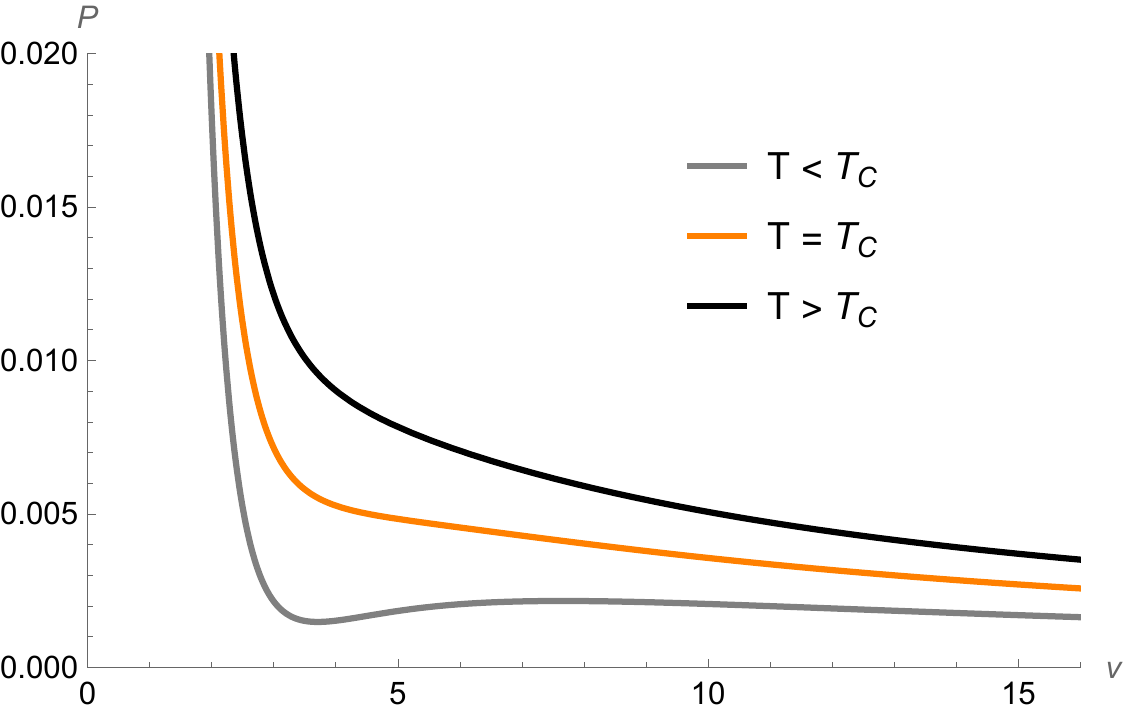}
	\caption{The behavior of $P$--$v$ isotherms for quintessence parameter $\omega_q=-2/3$, $\alpha=0.1$, $Q=1$, and varying thermodynamic temperatures $T$.}
	\label{isotherm}
\end{figure}

We now examine the critical behavior of a charged RN-AdS black holes with quintessence. Utilizing the expression of Hawking temperature  given in Eq.~(\ref{Hawking temp}), it is straightforward to obtain the thermodynamic equation of state expressed below as,
 \begin{equation}
 	P = \frac{T}{2r_+} - \frac{1}{8\pi r_+^2} + \frac{Q^2}{8\pi r_+^4} - \frac{3\alpha \omega_q}{8\pi r_+^{3(1+\omega_q)}}.\label{13}
 \end{equation}
In Eq.~(\ref{13}), the final term encodes the effect of the fundamental field, whereas the first three terms represent the usual equation of state for the RN-AdS black hole. Effectively lowering the pressure at a fixed horizon radius, the quintessence parameter $\alpha$ indicates the presence of a repulsive dark energy component that opposes the gravitational pull close to the event horizon. Introducing the specific volume $v = 2r_+$ \cite{Kubiznak:2012wp}, Eq.~(\ref{13}) transforms to
 \begin{equation}
 	P = \frac{T}{v} - \frac{1}{2\pi v^2} + \frac{2Q^2}{\pi v^4} - \frac{8^{\omega_q}\times 3\alpha \omega_q}{\pi v^{3(1+\omega_q)}}.\label{pv1}
 \end{equation}
In equation~(\ref{pv1}), the terms involving $1/v^{2}$ and $1/v^{4}$ reflect repulsive and attractive interactions between the microstructures, respectively, and are closely similar to the Van der Waals equation. In order to change the thermodynamic response and move the critical parameters based on the values of $\omega_q$ and $\alpha$, the quintessence correction adds an extra $v^{-3(1+\omega_q)}$ term \cite{Li:2014ixn,Ghosh:2017cuq}. The critical point is obtained from the inflection point of the $P$–$v$ isotherm, satisfying the following conditions:
 \begin{eqnarray}
 	\left(\frac{\partial P}{\partial v}\right)_{T=T_c} = 0, \quad 
 	\left(\frac{\partial^2 P}{\partial v^2}\right)_{T=T_c} = 0.\label{criticality condition}
 \end{eqnarray}
Applying these conditions to Eq.~(\ref{pv1}) yields the critical Hawking temperature and critical specific volume respectively as,
 \begin{equation}
 	T_c = \frac{1}{\pi v_c} - \frac{8Q^2}{\pi v_c^3} + \frac{8^{\omega_q}\times 9\alpha \omega_q(1+\omega_q)}{\pi v_c^{2+3\omega_q}}, \quad 	v_c^{2} - 24Q^2 + \frac{8^{\omega_q}(3\omega_q+2)(\omega_q+1)\times 9\alpha \omega_q}{v_c^{3\omega_q-1}} = 0
 \end{equation}
Substituting these into Eq.~(\ref{pv1}) provides the critical pressure,
 \begin{equation}
 	P_c = \frac{1}{2\pi v_c^2} - \frac{6Q^2}{\pi v_c^4} + \frac{8^{\omega_q}\times 3\alpha \omega_q(2+3\omega_q)}{\pi v_c^{3(1+\omega_q)}}.\label{102}
 \end{equation} 
For the particular case $\omega_q = -2/3$, the algebra simplifies, allowing analytic expressions for the critical quantities:
 \begin{equation}
 	v_c = 2\sqrt{6}Q, \quad
 	T_c = \frac{\sqrt{6}}{18\pi Q} - \frac{\alpha}{2\pi}, \quad
 	P_c = \frac{1}{96\pi Q^2}.\label{19}
 \end{equation}
The inclusion of the quintessential field fundamentally alters the critical temperature, adding a term proportional to $\alpha$, according to the analogy of Eq.~(\ref{19}) with the critical parameters of the RN-AdS black hole without quintessence \cite{Kubiznak:2012wp,GBCriticality}. This suggests that the phase transition is weakened by the dark energy field's tendency to lower the critical temperature. However, as the quintessential correction disappears at $\omega_q = -2/3$, the critical pressure and specific volume are unaffected. The $P$–$v$ isotherms corresponding to different temperatures are shown in Fig.~\ref{isotherm} by plotting Eq.~(\ref{pv1})
The isotherms show distinctive Van der Waals-like behavior with three separate branches for temperatures below the critical temperature $T_c$; small, intermediate, and large black holes. $\left(\partial P/\partial v \right)_T >0$, indicates that the intermediate branch is thermodynamically unstable~\cite{Li:2014ixn,Wei:2019uqg}, but the small and large branches correspond to locally stable configurations with positive specific heat. A first-order phase transition between small and giant black hole phases is shown by the behavior across the oscillation area. The physical interpretation of this behavior is that quintessence cools and dilutes the black hole system, hence lowering the temperature needed for phase coexistence while maintaining the general structure of the $P-v$ diagram. In addition, the presence of quintessence modifies the universality of the ratio $\frac{P_c v_c}{T_c}$, deviating from the usual Van der Waals behavior\footnote{See~\cite{Wei:2019uqg,Kumara:2020ucr,Zhou:2020vzf} and references therein for detail)}.
 %%%%%%%%%%%%%%%%%%%%%%%%
 
  \section{Thermodynamic criticality in the grand canonical ensemble}\label{Grand criticality}
The expression for the black hole mass as a function of the thermodynamic variables $P, V, \alpha$, and $\Phi$ can be rewritten in the grand canonical ensemble as follows,
 \begin{equation}\label{mass}
M = P V+\frac{1}{4} \sqrt[3]{V} \left(\sqrt[3]{\frac{6}{\pi }} \left(\Phi ^2+1\right)-2 \alpha  \sqrt{\left(\frac{16}{9}\right)^{\omega _q} \pi ^{2 \omega _q} V^{-2 \omega _q-\frac{2}{3}}}\right)
 \end{equation}
In order to satisfy the dominating energy requirement \cite{Hawking, Gibbons}, we set the crucial state parameter to \(\omega_q = -2/3\) in accordance with Kiselev \cite{Kiselev} throughout the remainder of the work.
Using the black hole's mass as enthalpy, the temperature may now be calculated once more as follows,
 \begin{equation}\label{temperature}
 	T = \sqrt[3]{\frac{6}{\pi }}PV^{1/3} - \frac{\alpha}{2\pi} - \frac{(\Phi^2-1)}{2\times 6^{1/3}\pi^{2/3}V^{2/3}}
 \end{equation}
The expression for the thermodynamic pressure $P$ as a function of $T$, $V$ (or specific volume $v$), $\alpha$, and $\Phi$ in the grand canonical ensemble can be obtained using Eqn.(\ref{Hawking temp}) as follows,
 \begin{eqnarray}\label{pressure GC}
 	P = \frac{12\pi TV^{1/3}+6V^{1/3}\alpha+6^{2/3}\pi^{1/3}(\Phi^2-1)}{12\times 6^{1/3}\pi^{2/3}V^{2/3}} = \frac{T}{v}+\frac{\alpha}{2\pi\,v}+\frac{\Phi^{2}-1}{2\pi\,v^{2}} ,
 \end{eqnarray}
where \(v\) denotes the specific volume, \(\alpha\) represents the quintessence normalization parameter and \(\Phi\) denotes the electric potential. The ideal-gas–like contribution is the first term, the quintessence-induced \(1/v\) correction is the \(\alpha\)-term, and the Van der Waals attractive/repulsive \(1/v^{2}\) contribution is the role of the \((\Phi^{2}-1)/v^{2}\) term (its sign determines whether interactions are effectively attractive or repulsive). 
 
 \subsection{Quenching of thermodynamic criticality within the grand canonical ensemble}

Starting with the equation of state for RN-AdS black holes with quintessence in the grand canonical ensemble, as given in Eq.~(\ref{pressure GC}),
\begin{eqnarray}
	P(v,T) =\frac{T}{v} + \frac{x}{v} + \frac{y}{v^{2}}, \label{eq:Pofv}\\
	x = \frac{\alpha}{2\pi}, \quad y=\frac{\Phi^{2}-1}{2\pi} \nonumber
\end{eqnarray}
and applying the critcality condition Eq.~(\ref{criticality condition}), we obtain the expressions for the first and second-order derivatives of pressure with respect to the specific volume v as follows,

\begin{eqnarray}
	\frac{\partial P}{\partial v} &=& -\frac{(T + x)}{v^{2}} - \frac{2y}{v^{3}}, \label{eq:firstderiv}\\
	\frac{\partial^{2} P}{\partial v^{2}} &=& \frac{2(T + x)}{v^{3}} + \frac{6y}{v^{4}}. \label{eq:secondderiv}
\end{eqnarray}
On solving these we get,
\begin{eqnarray}
	(T_c + x)v_c = 0. \label{eq:TcA}
\end{eqnarray}
Excluding the trivial unphysical case $v_c = 0$, the critical temperature is found to be
\begin{eqnarray}
	T_c = -x = -\frac{\alpha}{2\pi}. \label{eq:Tcfinal}
\end{eqnarray}
For physically meaningful quintessence fields, $\alpha > 0$, hence $T_c < 0$, which is thermodynamically unacceptable. Also, $y = 0$ is equivalent to $\Phi^2 = 1$, which is a specific, non-generic condition. The equation of state describing RN–AdS black holes with quintessence in the grand canonical ensemble does not admit any physical critical point $(T_c, P_c, v_c)$ for any physically justifiable choice of parameters. The characteristic inflection in the $P$-$v$ isotherms is generated by the presence of a higher order term ($v^{-4}$) in the equation of state in the traditional RN-AdS black hole, which results in a Van der Waals type first-order phase transition. However, in the current scenario, such higher order attractive factors are not present since the quintessence field dominates and truncates the equation of state. As a result, the critical point vanishes as the pressure monotonically drops as the specific volume increases. 

A new characteristic of the RN-AdS black hole with quintessence in the grand canonical ensemble is its apparent absence of criticality. In contrast to its canonical counterpart, it suggests that the system does not experience a small/large black hole phase transition. Rather, there is no coexistence region and the thermodynamic behavior stays smooth~\cite{Zhou:2020vzf}. These findings offer a crucial starting point for comprehending the microscopic interactions using thermodynamic geometry, which is covered in the subsequent section.
%%%%%%%%%%%%%%%%%%%%%%%%%%%%%%%%%%%%%%%%%%%%%%%%%%%%%%%%%%%
\section{Analysis of thermodynamic geometry}\label{Thermodynamic geometry}

Thermodynamic information geometry offers an efficient and generally accepted framework for analyzing the fundamental underlying microscopic degrees of freedom of black holes, especially in the context of the extended phase space perspective. The space of equilibrium thermodynamic states is transformed to a differentiable manifold with a Riemannian metric structure in this formalism, which was initially established by Ruppeiner~\cite{Ruppeiner,Ruppeiner2}. This geometric structure provides a covariant formulation of equilibrium fluctuation theory whereby the intrinsic geometry of the state space captures thermodynamic fluctuations~\cite{Janyszek:1989zz,Ruppeiner3,Janyszek1990,RuppeinerRMP,interactions}.

More precisely, a metric tensor defined as the Hessian of the entropy function \(S\) with respect to the appropriate extended variables captures the statistical interdependence among thermodynamic variables. As a result of this, the curvature measurements of this thermodynamic manifold directly indicate the particular kind of microscopic interactions which govern the system. The basic Boltzmann relation, leading to an association between entropy and the quantity of freely accessible microstates, serves as the foundation for such a framework as:
\begin{equation}\label{eq:microstates}
	\Omega = \exp\left( \frac{S}{k_B} \right),
\end{equation}
where  \( k_B \) denotes the usual Boltzmann constant and \( \Omega \) is the number of accessible microstates corresponding to a given macroscopic configuration. In order to investigate a fluctuations, a thermodynamic subsystem, with a set of independently extensive variables \(\xi^i\) (with \(i=1,2\)), is embedded in a much larger environment or reservoir \(\mathcal{\chi}_0\), so that the subsystem \(\mathcal{\chi} \subset \mathcal{\chi}_0\) exists in thermal equilibrium with its surroundings. In this configuration, the Gaussian approximation of fluctuation theory can handle slight deviations from equilibrium. Following that, we may express the probability distribution controlling these fluctuations near equilibrium as~\cite{RuppeinerRMP,interactions}
\begin{equation}\label{Pdx}
	\mathbb{P}(\xi^1, \xi^2) \propto \exp\left( -\frac{1}{2} \delta l^2 \right),
\end{equation}
where the thermodynamic line element is defined by the exponent using a quadratic form.

The quantity \( \delta l^2 \) denotes the distance between neighboring equilibrium states in the thermodynamic state space and is expressed as,
\begin{equation}\label{eq:distance}
	\delta l^2 = -\frac{1}{k_B} \frac{\partial^2 S}{\partial \xi^i \partial \xi^j} \, \delta \xi^i \delta \xi^j.
\end{equation}
This expression indicates the system's sensitivity to small fluctuations in the extensive variables and defines the Ruppeiner metric in terms of the second derivatives of entropy. In geometrical terms, Eq.~\ref{eq:distance}) provides an exact measure of distinguishability between neighboring macrostates by acting as the infinitesimal line element on the manifold of thermodynamic states.

Furthermore, this geometric structure has significant physical implications in addition to being merely formal: the thermodynamic distance's magnitude controls the probability of fluctuations, and the associated curvature invariants are known to provide information regarding the type and strength of the system's effective microscopic interactions. Equation~(\ref{Pdx}) shows that the fluctuation probability strictly increases with the closeness of any two points on the space\cite{RuppeinerRMP,interactions}. Thus, the metric \(g_{ij} = -\partial_i \partial_j S\), which forms the foundation of thermodynamic geometry, encodes the Gaussian fluctuations around equilibrium. A crucial indicator of the underlying microscopic interaction structure, the Ricci scalar \(R\), which is derived from the thermodynamic metric tensor \(g_{ij}\), describes the type and strength of correlations between the constituent degrees of freedom. On the other hand, the thermodynamic line element's magnitude indicates the probability of spontaneous fluctuations; higher transition probabilities between neighboring equilibrium states are associated with smaller distances. Additionally, there is a direct physical explanation for the sign of the scalar curvature \(R\): a negative value is generally understood to indicate attractive interactions among the microscopic elements, whereas a positive value is generally linked with effectively repulsive interactions. Thus, the the behavior of \(R\) offers significant details on the statistical mechanical properties of the system~\cite{Wei:2019yvs,Wei:2015iwa}. In systems where \(R = 0\) there is no interaction. Specifically, divergences in \(R\) frequently indicate phase transitions since they are accompanied by diverging susceptibilities like compressibility or specific heat capacity. Furthermore, the correlation length \(\zeta\)  confronting criticality is given by \( |R| \sim \zeta^{\tilde{d}} \), where \( \tilde{d} \) is the spatial dimensionality of the system.

Within the framework of four-dimensional anti-de Sitter (AdS) black hole thermodynamics, the application of information geometry proves to be particularly powerful for analyzing critical phenomena, especially for electrically charged configurations studied under fixed thermodynamic ensembles. The Ruppeiner scalar curvature has been demonstrated in several recent investigations~\cite{Singh:2023ufh,Wei:2019yvs,Wei:2015iwa} to act as a sensitive probe of phase structure, successfully capturing signatures of both first-order and second-order phase transitions. These include, notably, the Hawking--Page transition as well as phase behavior analogous to the van der Waals liquid--gas system.

More specifically, for charged AdS black holes, the behavior of the scalar curvature \( R \) depends intricately on the electric charge sector and the choice of thermodynamic ensemble employed in the analysis. It is observed that divergences in \( R \) occur in correspondence with singular points of response functions such as the specific heat or isothermal compressibility. Such divergences therefore serve as clear indicators of thermodynamic instability and criticality within the system. Furthermore, zero-crossings or sign reversals in $R$, which correspond to phenomena seen in quantum many-body systems and Bose-Fermi gases, propose crossovers between interaction-dominated microstructures. Although it is not immediately clear, these findings support the increasing notion that black hole microstructure can be described by thermodynamic geometry. Adopting this geometric framework, we analyze electrically charged AdS black holes within the grand canonical ensemble, with the objective of examining how the Ruppeiner scalar \( R \) evolves across the relevant parameter space. This investigation enables us to extract detailed information encoded in \( R \), thereby shedding light on the intrinsic organization and interaction characteristics of the underlying quantum gravitational microstates.

%%%%%%%%%%%%%%%%%%%%%%%%%%%%%%%%%%%%%%%%%%%%%%%%
%%%%%%%%%%%%%%%%%%%%%%%%%%%%%%%%%%%%%%%%%%%%%%%%%%%
 
% % % % % % % % % % % %
\subsection{Behavior of thermodynamic geometry  in the grand canonical ensemble}\label{Ruppeiner geometry}
For any arbitrary thermodynamic potential expressed as \(\phi = \phi(\xi^i)\), the corresponding metric is defined by,
\begin{equation}
  (\delta l)^2 = \frac{\partial^2 \phi}{\partial \xi^i \partial u\xi^j} d\xi^i d\xi^j.
\end{equation}
It is possible to think of this design as defining an effective distance measure between nearby equilibrium thermodynamic configurations. An alternate pair of independent fluctuation variables must be introduced because, in the current situation, the variables \(S\) and \(V\) do not form an independent set. It is possible to formulate the relevant metric structure in terms of the Hessian of the enthalpy for static black holes thanks to the natural choice offered by \(S\) and \(P\). It is crucial to note that the probabilistic description of thermodynamic fluctuations in fluctuation theory is inextricably linked to a distance measure that is obtained from the negative Hessian of the entropy. The definition of a thermodynamic length is motivated by this underlying link. In contrast, the length constructed from a Weinhold-type metric does not, in general, admit a direct physical interpretation in this probabilistic framework. 

\begin{figure}[h!]
  	\begin{center}
  		\centering
  		\includegraphics[width=4.0in]{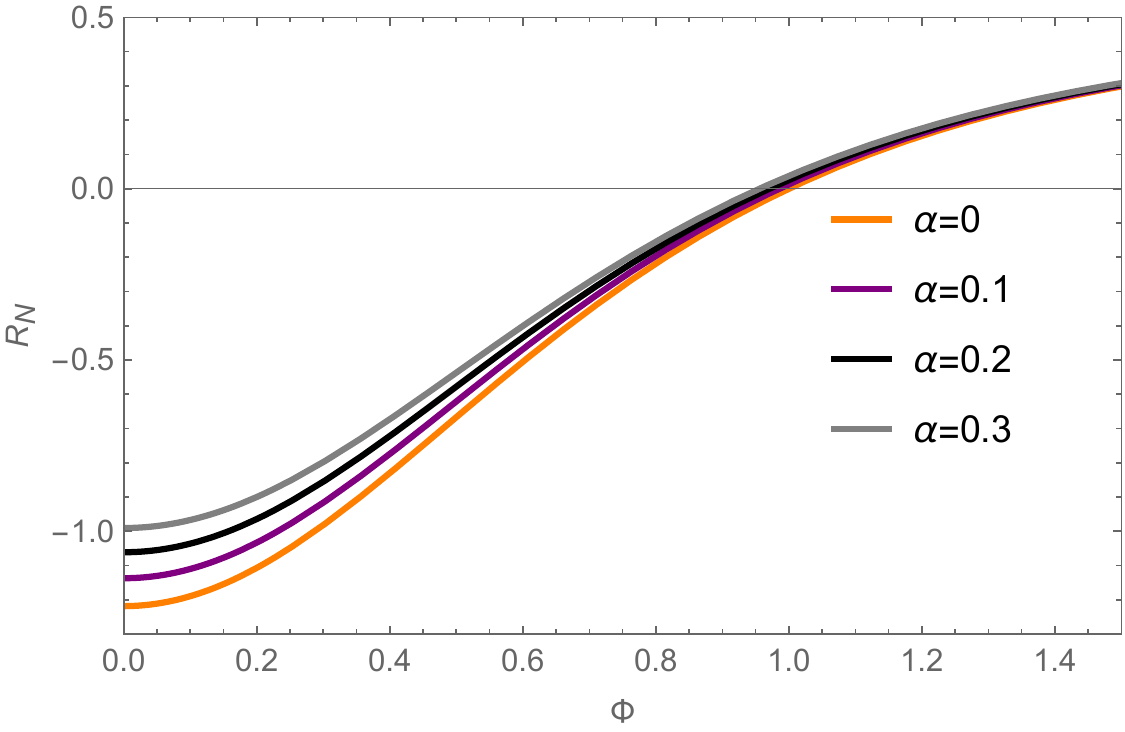}  		
  		\caption{The variation of the normalized Ruppeiner curvature scalar \(R_N\) with respect to the electric potential \(\Phi\), for fixed values of \(V=0.1\), \(T=1.2\), and different choices of the parameter \(\alpha\) in the grand canonical ensemble. The plot highlights the transition in the nature of microscopic interactions as \(\Phi\) varies.}
          \label{Rn_phi}	
  	\end{center}
  \end{figure}

Nevertheless, it is well understood that a suitable conformal transformation restores its physical relevance: specifically, rescaling the Weinhold-like metric by a factor of \(1/T\) yields a quantity that is consistent with the fluctuation-theoretic interpretation. This establishes a correspondence with the Ruppeiner metric, as discussed previously.
%%%%%%%%%%%%%%%%%%%%
We now proceed to evaluate the Ruppeiner curvature on the \((T,V)\)-plane, which provides a more appropriate thermodynamic representation for analyzing black hole microstructures in the extended phase space. This choice is particularly motivated by the fact that, for a large class of black hole solutions, the entropy \(S\) and thermodynamic volume \(V\) are not independent fluctuation variables\footnote{As discussed in \cite{Singh:2020tkf}, the \((S,V)\)-plane becomes unsuitable for several black hole systems due to the functional dependence between the fluctuation coordinates \(S\) and \(V\).}. Therefore, one must instead employ alternative thermodynamic potentials, such as enthalpy or Helmholtz free energy, to construct a consistent fluctuation geometry. In this representation, the Ruppeiner line element on the \((T,V)\)-plane is expressed as\cite{Xu,Ghosh,Singh:2020tkf},
\begin{equation}\label{RuppeinerTV}
  dl_{R}^2 =  \frac{C_V}{T^2}dT^2 +  \frac{1}{T}\bigg(\frac{\partial P}{\partial V}\bigg)_TdV^2
\end{equation}
Here, \(C_V\) represents the specific heat at constant volume. For the black hole system under consideration, this quantity identically vanishes\footnote{This follows from the fact that the horizon entropy is determined entirely by the geometric volume, leading to no independent variation of entropy with temperature at fixed volume.}. Moreover, the geometric volume is exactly equivalent to the thermodynamic volume in this setup. This property is generally valid for static (non-rotating) black holes in four or higher spacetime dimensions. By making use of Eq.~(\ref{pressure GC}), Eq.~(\ref{entropy}), and the line element given in Eq.~(\ref{RuppeinerTV}), one can explicitly evaluate the normalized Ruppeiner scalar curvature. The resulting expression is given by,
\begin{equation}\label{Rn}
R_N = \frac{\left(3V^{1/3}\alpha+6^{2/3}\pi^{1/3}(\Phi^2-1)\right)\left(12\pi TV^{1/3}+3V^{1/3}\alpha +6^{2/3}\pi^{1/3}(\Phi^2-1) \right)}{2\left(6\pi TV^{1/3}+3V^{1/3}\alpha+6^{2/3}\pi^{1/3}(\Phi^2-1)\right)^2}
\end{equation}
The normalized Ruppeiner curvature scalar serves as a diagnostic tool for understanding the nature of interactions among the underlying microscopic degrees of freedom of the black hole system. In particular, the sign of \(R_N\) provides information about whether the dominant interactions are attractive or repulsive. 
Additionally, we investigate how the electric potential affects the curvature scalar in order to examine the microstructural behavior in the grand canonical ensemble\cite{Zhou:2020vzf}. It is clear from Eqn.~(\ref{Rn}) that \(R_N\) depends directly on \(\Phi^2\), suggesting that the system's interaction features are significantly influenced by the electric potential's amplitude. We now investigate the behavior of the normalized Ruppeiner curvature scalar \(R_N\). For fixed values of \(T\) and \(\alpha\), Fig. (\ref{Rn_phi}) illustrates the variation of \(R_N\) as a function of the electric potential \(\Phi\).
From the behavior shown in Fig. (\ref{Rn_phi}), three key observations can be made:
\begin{itemize}
\item First, the normalized Ruppeiner curvature scalar vanishes at $\Phi = \sqrt{1 -\frac{\alpha}{2} \sqrt[3]{\frac{6V}{\pi }}},$ where all the curves corresponding to different parameter values intersect. At this point, the curvature becomes zero, indicating that the effective interaction between the microscopic constituents disappears. This behavior resembles that of an ideal gas system, and notably, it appears to be independent of the thermodynamic parameters \(T\), \(V\), and \(\alpha\), suggesting a universal feature.

\item In the range $0<\Phi< \sqrt{1 -\frac{\alpha}{2} \sqrt[3]{\frac{6V}{\pi }}}$, the curvature scalar takes negative values, i.e., \(R_N<0\). This indicates that the dominant interaction among the black hole microstructures is attractive in this regime. Additionally, the magnitude \(|R_N|\) is found to increase with increasing thermodynamic volume \(V\), implying that the strength of attractive interactions becomes more pronounced for larger volumes.

\item For $\Phi>\sqrt{1 -\frac{\alpha}{2} \sqrt[3]{\frac{6V}{\pi }}}$, the curvature scalar becomes positive, \(R_N>0\), which signifies the dominance of repulsive interactions. In this region, a decrease in volume leads to an increase in \(R_N\), indicating stronger repulsive effects. It is also observed that both the critical temperature and critical pressure assume negative values in this parameter range, which implies that the conventional small-large black hole phase transition does not occur.
\end{itemize}

We next analyze the microstructural properties of the charged AdS black hole in the grand canonical ensemble by examining the behavior of \(R_N\) in the \(T-V\) phase diagram. In particular, we introduce the sign-changing curve\cite{Zhou:2020vzf}, defined by the condition \(R_N = 0\), along with the spinodal curve determined by $\left(\frac{\partial P}{\partial V}\right)_T = 0$.
\begin{figure}[h!]
  	\begin{center}
  		\centering
  	\includegraphics[width=3.5in]{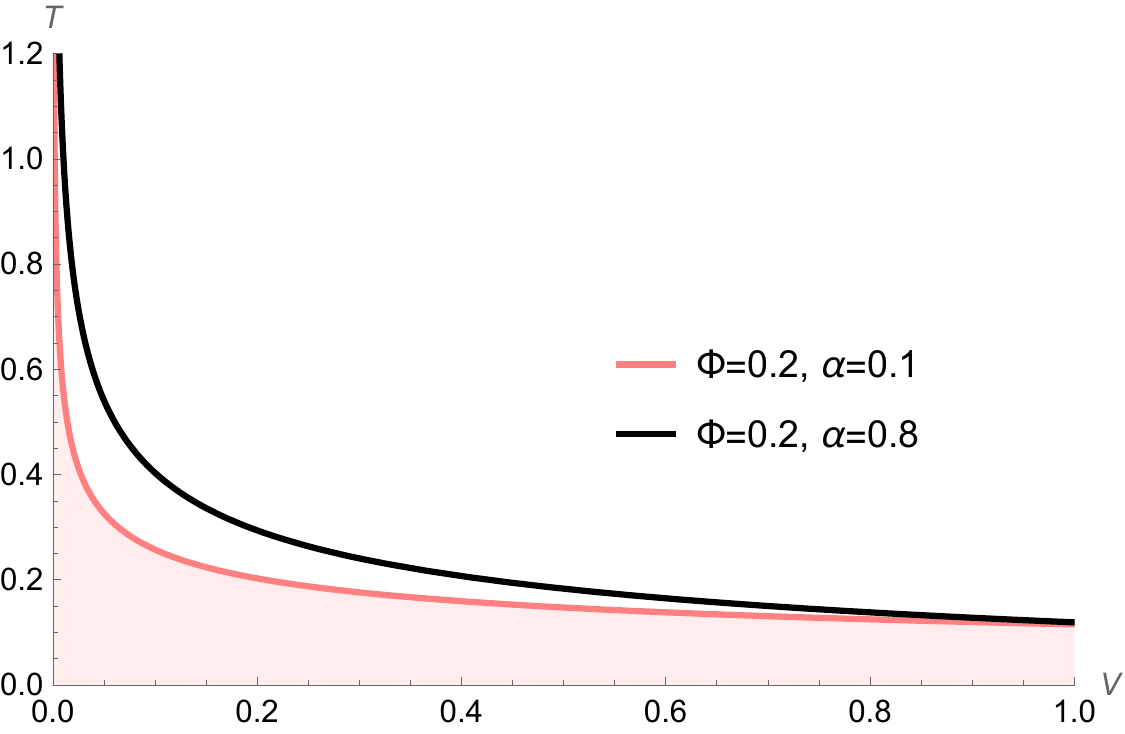}  		
  		\caption{The spinodal curve (black) and the sign-changing curve (pink) plotted in the \(T-V\) plane for fixed values of \(\Phi\) and \(\alpha\). The shaded region corresponds to \(R_N >0\), indicating repulsive interactions, while the remaining region corresponds to \(R_N <0\), indicating attractive interactions.}
          \label{spinodal}	
  	\end{center}
  \end{figure}
The normalized Ruppeiner curvature exhibits divergence along the spinodal curve, which is given by,
\begin{equation}
T_{div}= \frac{6^{2/3}\pi^{1/3}(1-\Phi^2) - 3V^{1/3}\alpha}{6\pi V^{1/3}}
\end{equation}
The sign-changing curve divides the \(T-V\) plane into regions of positive and negative curvature and is defined as,
\begin{equation}
T_0 = \frac{T_{div}}{2}=  \frac{6^{2/3}\pi^{1/3}(1-\Phi^2) - 3V^{1/3}\alpha}{12\pi V^{1/3}}
\end{equation}
From the above relations, it follows that the condition $T_{div}=2T_0$ holds, which appears to be a general feature across different black hole systems. As can be inferred from Fig.~(\ref{spinodal}), the region below the sign-changing curve corresponds to \(R_N>0\), whereas the region above it corresponds to \(R_N<0\). This indicates that the dominant microscopic interaction is primarily attractive over a significant portion of the parameter space, consistent with the behavior observed in charged four-dimensional AdS black holes. It is therefore evident that, although the electric potential and charge influence the thermodynamic properties, they do not qualitatively alter the nature of the dominant interaction in most regions. However, for $\Phi> \sqrt{1 -\frac{\alpha}{2} \sqrt[3]{\frac{6V}{\pi }}}$, the disappearance of the small-large black hole phase transition indicates a substantial modification in the thermodynamic behavior of the system.

\begin{figure}[h!]
  	\begin{center}
  		\centering
  	\includegraphics[width=3.5in]{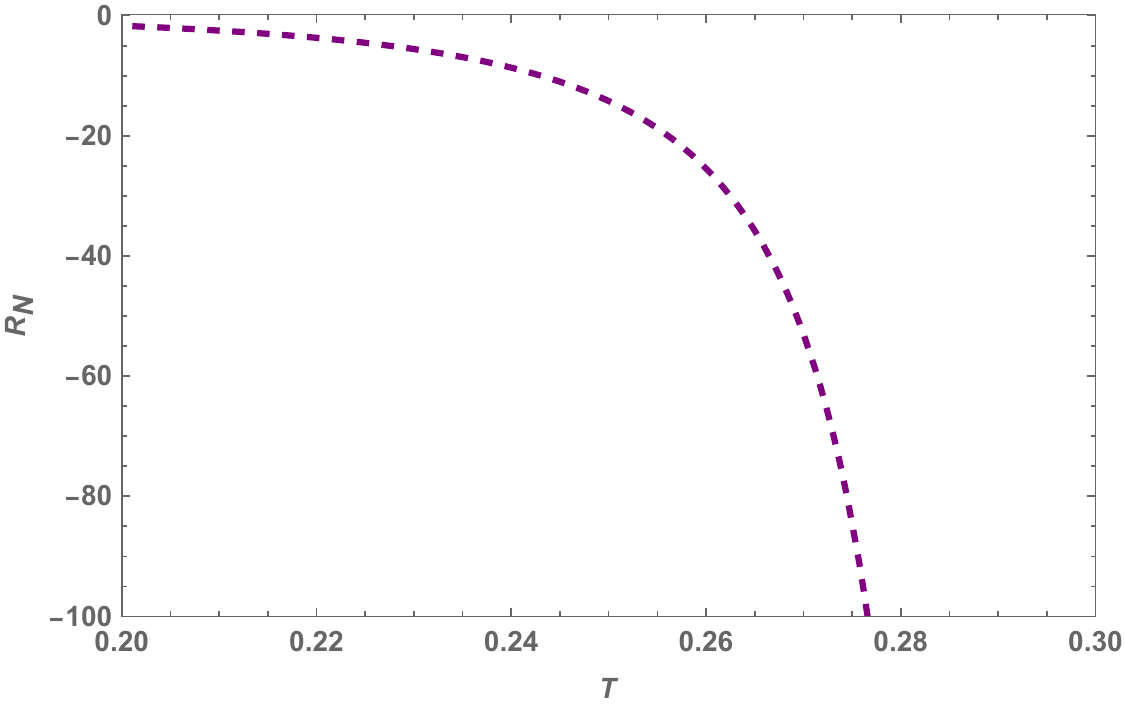}  		
  		\caption{The variation of the normalized Ruppeiner curvature scalar \(R_N\) as a function of temperature \(T\), for fixed values of \(V=0.24\), \(\Phi=0.3\), and \(\alpha=0.5\). The plot demonstrates the evolution of interaction strength as the system approaches the critical temperature.}
          \label{RN_vs_T}	
  	\end{center}
  \end{figure}

Finally, we examine the variation of the normalized Ruppeiner curvature scalar with temperature, as shown in Fig.~\ref{RN_vs_T}.
As the temperature increases, the curvature scalar initially takes negative values and gradually decreases further. This trend continues until \(R_N\) diverges to negative infinity at the critical temperature. This divergence signals the onset of critical behavior characterized by strong fluctuations. Notably, the curvature remains negative throughout, indicating that the dominant interaction remains attractive even as the system undergoes significant changes in its phase structure.

%%%%%%%%%%%%%%%%%%%%%%%%%%%%%%%%%%%%%%%%%%%%%%%%%%%%%%%%%%%%

\section{Remarks}\label{Remarks}
The significant properties of microstructures for RN-AdS black holes with quintessence in the grand canonical ensemble have been phenomenologically investigated in the current study. First, we examined the $P-v$ criticality and constructed the charged AdS black hole's thermodynamic properties using quintessence. We noticed that the quintessence in the form of dark energy caused an additional word to appear in the critical temperature equation. We also looked at how the $P-v$ isotherms behaved. The aforementioned research shows that the thermodynamic equation of state loses its distinctive Van der Waals structure when the RN-AdS black hole is investigated in the grand canonical ensemble and with quintessence.

The resulting absence of $P-v$ criticality is a significant and novel result, providing a deeper insight into how quintessence modifies both the macroscopic and microscopic thermodynamic behavior of black holes. The Ruppeiner thermodynamic geometry for RN-AdS black holes with quintessence in the grand canonical ensemble was then calculated analytically. We also calculated the equivalent normalized Ruppeiner curvature scalar and observed some remarkable and distinctive features. For lower electric potential $\Phi$ values, $R_N$ turns out to be negative, indicating that attractive interaction between the black hole microstructures predominates. It is observed that the effective interaction becomes repulsive only when the electric potential exceeds a threshold value, namely \(\Phi > \sqrt{1 -\frac{\alpha}{2} \sqrt[3]{\frac{6V}{\pi }}}\,.\) In addition, we have analyzed the structure of the spinodal curve and the corresponding sign-changing curve, together with the behavior of the Ruppeiner scalar curvature in the \(T\!-\!V\) phase plane. The critical properties of the normalized Ruppeiner scalar, \( R_N \), were also investigated in detail. In particular, it is found that \( R_N \) diverges to negative infinity as the system approaches the critical point, signaling the onset of critical behavior. By combining these observations, we were able to systematically classify the nature of microscopic interactions across the parameter space. 

The resulting behavior differs from that of neutral Gauss--Bonnet AdS black holes, while exhibiting close agreement with the characteristics observed for electrically charged AdS black holes. These findings provide important insight into the underlying microstructural properties of black holes in the grand canonical ensemble. Furthermore, this analysis can be naturally extended to more general scenarios, including rotating configurations as well as higher-dimensional neutral and charged Gauss--Bonnet AdS black holes.

%%%%%%%%%%%%%%%%%%%%%%%%%%%%%%%%%%%%%%%%%%%%%%%%%%%%%%%%%%%
% % % % % % % % % % % % % % % %
\section{Acknowledgements}

The authors extend their appreciation to the Deanship of Research and Graduate Studies at King Khalid University for funding this work through Small Research Project under grant number RGP1/107/46.

\section*{Data Availability Statement}
Data sharing is not applicable to this article since no datasets were generated or analyzed during the present study.

\section*{Conflict of Interests}
The author declares that there are no competing financial or non-financial interests in relation to this work.
% % % % % % % % % % % % % % %
% % % % % % % % % % % % % % % % %

\end{document}